\documentclass{webofc}
\usepackage[varg]{txfonts}   
\usepackage{gensymb}
\begin{document}
\title{QCD at Finite Temperature and Density - Equation of State}
%
%

\author{\firstname{Jamie} \lastname{Karthein}\inst{1}\thanks{\email{jmkar@mit.edu}} 
}

\institute{Center for Theoretical Physics, Massachusetts Institute of Technology, Cambridge, MA 02139, USA 
          }

\abstract{
As an important set of thermodynamic quantities, knowledge of the equation of state over a broad range of temperatures and chemical potentials in the QCD phase diagram is crucial for our understanding of strongly-interacting matter.
There is a good understanding from first-principles results in lattice QCD, perturbative QCD and chiral effective field theory about the equation of state.
However, these approaches are valid in different regimes of the phase diagram, and therefore, a method of providing an equation of state that covers a full range of the phase diagram involves matching together these results with appropriate models in order to fill in the gaps between these regions.
Furthermore, with such equations of state, important questions about QCD phase structure can begin to be addressed, such as whether there is a critical point in the QCD phase diagram.
In this contribution to the proceedings, equations of state from first-principles and effective theories will be discussed in order to understand how QCD thermodynamics is affected by the presence of a critical point.}
\maketitle
\section{Introduction}
\label{intro}
The phase structure of strongly-interacting matter has been a focus of efforts for both the theoretical and experimental communities for decades. Upon exploration of the QCD phase diagram, the question being probed is how does strongly-interacting matter respond to heat, i.e. increasing temperature, and compression, i.e. increasing density. Therefore, the QCD equation of state (EoS) is the object of study lending from the fact that these are thermodynamic properties. First-priniciples lattice QCD calculations have given insight into the zero and low-to-moderate density equation of state \cite{Borsanyi:2010cj,Bazavov:2014pvz,Bazavov:2017dus,Borsanyi:2021sxv,Mondal:2021jxk}. Also from the lattice with good precision, the transition temperature at vanishing chemical potential has been calculated as well as the curvature of the transition line and where it lies up to $\mu_B \simeq 2.5T$ \cite{Borsanyi:2020fev,Aarts:2023vsf}. In the low-temperature region of the QCD phase diagram, the system is rather well-described by an ideal gas of hadrons and their resonances \cite{Borsanyi:2010bp,Vovchenko:2016rkn,Alba:2020jir, SanMartin:2023zhv}. However, as the density, or equivalently chemical potential, increases, nuclear matter exhibits a liquid-gas phase transition in the low temperature regime that is described by an interacting van der Waals equation of state \cite{Vovchenko:2016rkn}. Furthermore, chiral effective field theory is the appropriate low energy theory that covers the low-temperature and low-to-moderate-density regime \cite{Drischler:2021kxf}. On the other hand, at asymptotically high temperatures and densities, perturbative QCD (pQCD) results yield access to the asymptotic regime of the QCD phase diagram \cite{Andersen:2002jz,Kurkela:2009gj,Haque:2013sja}. From the perspective of connecting to the observable universe, zero temperature calculations from chiral effective field theory and pQCD can be combined with astrophysical constraints to learn about the QCD equation of state \cite{Annala:2019puf,Mroczek:2023zxo}. On the other hand in terms of terrestrial experiments, along the $T-$axis of the QCD phase diagram are high energy heavy-ion collisions, such as those conducted at the Large Hadron Collider (LHC) at CERN \cite{ALICE:2022wpn,Busza:2018rrf}. 
Furthermore, by decreasing collider energy, the phase diagram at non-zero baryon chemical potential can be explored, which is a main focus of the Beam Energy Scan (BES) program at the Relativistic Heavy Ion Collider (RHIC) at Brookhaven National Lab \cite{Bzdak:2019pkr}. 

In light of these varied sources across theory, experiment, and astrophysical observation, there is a plethora of information about the phase structure of QCD within the regimes of validity of each of them \cite{MUSES:2023hyz}. 
These, therefore, yield the ground truth knowledge of the QCD EoS, which, however, does not cover the entire reach of the phase diagram. Thus, many questions remain about the phases of QCD. For example, does the crossover phase transitions proven by lattice calculations \cite{Aoki:2006we} become a proper first order phase transition at larger chemical potentials? This would imply that there is a critical point somewhere in the phase diagram where the change in transition type occurs. The search for this critical point has been pursued in part by studying critical fluctuations related to the diverging correlation length at the critical point \cite{Stephanov:1998dy,Stephanov:1999zu,Stephanov:2011pb}. On the other hand, where does the transition line lie at high densities which are beyond the current reach of lattice calculations? This question in essence probes whether the densities found in neutron stars are large enough to create deconfined quark matter. There are a number of further profound questions about the nature of strongly-interacting matter beyond the reaches of current experimental and theoretical methods. Thus, in order to explore these questions from the theory side, a method of piecing the current first-principles knowledge together with models of QCD is required in order to fill in the gaps in the phase diagram. The goal of the MUSES collaboration is to provide equations of state along these lines. The forthcoming publicly available software will allow for user-specified coverage of the phase diagram and user-defined conditions in addition to the choice of model in the region of applicability.

As mentioned previously there are many open questions about QCD phase structure,  here I will focus on the regime covered by high energy heavy-ion collisions due to the brevity required for proceedings. 
In this contribution, I will address the question of how QCD thermodynamics are affected by critical features in the phase diagram.

\section{Equation of State at Finite Temperature and Density}
\label{eos}


\subsection{Utilizing a critical equation of state to estimate equilibrium proton fluctuations}
\label{critical-EoS}
In order to study the effect of a critical point that could potentially be observed in the BES-II program at RHIC, the 3D Ising model is utilized to map such critical behavior onto the phase diagram of QCD.
The 3D Ising model was chosen for this approach because it exhibits the same scaling features in the vicinity of a critical point as QCD, in other words, they belong to the same universality class \cite{Pisarski:1983ms,Rajagopal:1992qz}.
The mapping of the critical behavior onto the QCD phase diagram is, however, not universal, which is to say that there are no strict mapping parameters that exist \textit{a priori} between the Ising phase diagram and the one for QCD.
Such a map can be constructed in the following way as in the BEST equation of state \cite{Parotto:2018pwx, Karthein:2021nxe}:
    \begin{equation} \label{mapT}
        \frac{T-T_c}{T_c}=\omega(\rho r \sin{\alpha_1} + h \sin{\alpha_2})
    \end{equation}
    \begin{equation} \label{mapmuB}
        \frac{\mu_B-\mu_{B,c}}{T_c} = \omega(-\rho r \cos{\alpha_1} - h \cos{\alpha_2})
    \end{equation}

Utilizing this mapping, we study the critical fluctuations in the Ising model in order to calculate the proton fluctuations near a critical point. We provide an update to the approach laid out in Ref. \cite{Athanasiou:2010kw} for calculating such particle fluctuations. This update includes utilizing the critical correlation length from universality. We, additionally, further rely on universality in order to extract the higher order couplings, $\lambda$, of the fluctuations:
\begin{equation} \label{eq:kappas-lambdas}
    \begin{split}
        \kappa_3 = 2 \lambda_3 V T^2 \xi^6 \qquad\qquad
        \kappa_4 = 6 V T^3 [2 (\lambda_3 \xi)^2 - \lambda_4] \xi^8
    \end{split}
\end{equation}

Following the work of Pradeep, Rajagopal, Stephanov, and Yin on Gaussian fluctuations \cite{Pradeep:2022mkf}, we explore the higher order non-Gaussian fluctuations in- and out-of-equilibrium. As pertains to this contribution centered on the equation of state, an equilibrium thermodynamic property, we will focus on the equilibrium results related to that work.

The first part of updating the estimates for proton fluctuations involves the input for the correlation length from universality. The scaling form of the correlation length in the Ising model, which follows Widom’s scaling form in terms of Ising model variables, can be written as \cite{ZinnJustin}: 
\begin{equation} \label{corr_length}
    \begin{split}
        \xi^2(r,M) = M^{-2 \nu / \beta} f_\xi(r/M^{1/\beta}),
    \end{split}
\end{equation}
where $\nu$ is the correlation length critical exponent, $f_\xi$ is the scaling function and the scaling parameter is $x$=$\frac{r}{M^{1/\beta}}$. 

In the $\epsilon$-expansion, the function $f_\xi$ is given to $\mathcal{O}$($\epsilon^2$) as \cite{BREZIN1974285}:
\begin{equation} \label{eq:g_epsilonexp}
    \begin{split}
        f_\xi(x) &= (f^+_1)^2 6^{-2 \nu} z \Big\{ 1 - \frac{\epsilon}{36} [(5 + 6 \ln 3)z - 6(1+z) \ln z ] 
        + \epsilon^2 \Big[\frac{1+2z^2}{72} \ln^2 z + \Big(\frac{z}{18} \Big(z - \frac{1}{2}\Big) (1 - \ln 3) \\
        &- \frac{1}{216} \Big(16z^2 - \frac{47}{3} z - \frac{56}{3} \Big) \Big) \ln z 
        + \frac{1}{216} \Big( \frac{101}{6} + \frac{2}{3}I + 6\ln^2 3 + 4 \ln 3 - 10 \Big) z^2 \\
        &- \frac{1}{216} \Big( 6 \ln^2 3 + \frac{44}{3} \ln 3 + \frac{137}{9} + \frac{8}{3}I \Big) z \Big] \Big\}
    \end{split}
\end{equation}
where $f^+_1$ is a non-universal critical amplitude, $z \equiv \frac{2}{1+x/3}$, $I \equiv \int_0^1 \frac{\ln [x (1-x)]}{1 - x(1-x)}dx \sim -2.3439$, and $\epsilon = 4 - d$, where $d$ is the spatial dimension.

One may also write the parametric form of the correlation length in the Ising model in terms of the parametric variables $(R,\theta)$ as \cite{ZinnJustin}:
\begin{equation}
\label{ZinnJustin}
    \xi^2(M,t) = R^{-2 \nu} g_\xi(\theta), \quad
    g_\xi(\theta) = g_\xi(0)(1-\frac{5}{18}\epsilon \theta^2 + O(\epsilon^2)).
\end{equation}
In this case the parametric equation of state is defined as:
\begin{equation} \label{IsingEoS}
        \begin{split}
            M = M_0 R^{\beta} \theta \qquad
            h = h_0 R^{\beta \delta} \tilde{h}(\theta) \qquad
            r = x_0 R(1- \theta^2)
        \end{split}
    \end{equation}
where   $x_0 = b^{1/\beta}/(b^2-1), \,
        h(\theta) = h_0 \theta (b^2-\theta^2)(1+c\theta^2) +\mathcal{O}(\epsilon^4),\, 
        b^2=\frac{3}{2}(1-\frac{\epsilon^2}{12}),\, 
        c=-\frac{\epsilon^3}{18}(\zeta(3)+\frac{I-1}{4})$.
%
With these calculations, we further expand the correlation length to $\mathcal{O}(\epsilon^2)$.
Thus, the $\theta$-dependence of the correlation length to $\mathcal{O}(\epsilon^2)$ is then given by:
\begin{equation}
\label{xiOeps2}
    \tilde{g}_\xi = \tilde{g}_\xi (0)\Bigg(1-\frac{5}{18}\epsilon \theta^2 + \Big[\frac{1}{972}( 24 I-25)\theta^2 + \frac{1}{324} ( 4 I + 41) \theta^4\Big] \epsilon^2)\Bigg) .
\end{equation}
%
We utilize this updated correlation length to $\mathcal{O}(\epsilon^2)$ in order to be consistent with the equation of state which is valid to the same order in the $\epsilon$-expansion.

On the other hand, we also take into account the $\mu_B$-dependence of the couplings $\lambda_3$ and $\lambda_4$, as shown in Eq. \ref{eq:kappas-lambdas}. Since we know that the 4th order moment is expected to exhibit a sign change \cite{Stephanov:2011pb}, these couplings will not simply be constants, which provides another modernization of the work in Ref. \cite{Athanasiou:2010kw}. We extract these higher order couplings via the knowledge of the correlation length and fluctuations in the Ising model. In order to study this $\mu_B$-dependence of the critical correlation, we need to choose the non-universal parameters of the Ising-QCD mapping. We choose parameters for our EoS in order to be consistent with recent results from STAR \cite{STAR:2020tga}.
Furthermore, our choice also obeys constraints from universality studied by Pradeep and Stephanov in Ref. \cite{Pradeep:2019ccv} as well as those given by causality and stability of the BEST EoS as shown by Mroczek et al in Ref. \cite{Mroczek:2022oga}.
As such, we move away from the choice of orthogonal Ising axes as given by the canonical choice in the BEST EoS in Refs. \cite{Parotto:2018pwx,Karthein:2021nxe} in order to agree with constraints from the universal features of this mapping shown by Pradeep and Stephanov.
Consequently, in order to have both the constraints properly taken into account and a strength of divergence that will reasonably agree with results from STAR net-proton energy dependence, we must modify the choice of scaling parameters, $w, \rho$.
However, as shown in the analysis of the BEST EoS parameter space by Mroczek et al in Ref. \cite{Mroczek:2022oga}, decreasing the angle difference, i.e. moving away from orthogonality, leads to a restriction of the allowed values for $w, \rho$.
Thus, we arrive at the following choice of mapping parameters.
\begin{center}
   \begin{tabular}{c c c c}
        $\mu_{B,c} = 420$ MeV & $\Delta \alpha =\alpha_1 - \alpha_2= 10 \degree$ &
        $w=8$ & $\rho=0.2$
   \end{tabular}
\end{center}

Finally, with inputs from universality for both the correlation length and the $\mu_B$-dependent couplings, we calculate the 3rd and 4th order proton fluctuations, similar to Ref. \cite{Athanasiou:2010kw},
\begin{equation}
\omega_{3p,\sigma}= \frac{2\tilde{\lambda}_3}{T^{3/2} \, n_p} \xi^{9/2} \Bigg( d_p \, g_p \int_k \frac{v_k^2}{\gamma_k} \Bigg)^3 \qquad
\omega_{4p,\sigma}= \frac{6(2\tilde{\lambda}_3^2 - \tilde{\lambda_4})}{T^2 \, n_p} \xi^7 \Bigg( d_p \, g_p \int_k \frac{v_k^2}{\gamma_k} \Bigg)^4
\label{eq:omegas}
\end{equation}
where the dimensionless $\lambda$'s are given by $\tilde{\lambda}_3=\lambda_3 (T^{1/2} \xi^{3/2})$, $\tilde{\lambda}_4=\lambda_4 (T \xi)$, $n_p$ is the proton density, and the kinematic integrals involve the degeneracy, $d_p$, and the proton coupling to the sigma field $g_p$. 
We consider a coupling of $g_p\sim7$ as in Ref. \cite{Athanasiou:2010kw}.

\begin{figure*}
\centering
\includegraphics[width=0.45\textwidth, height=5cm]{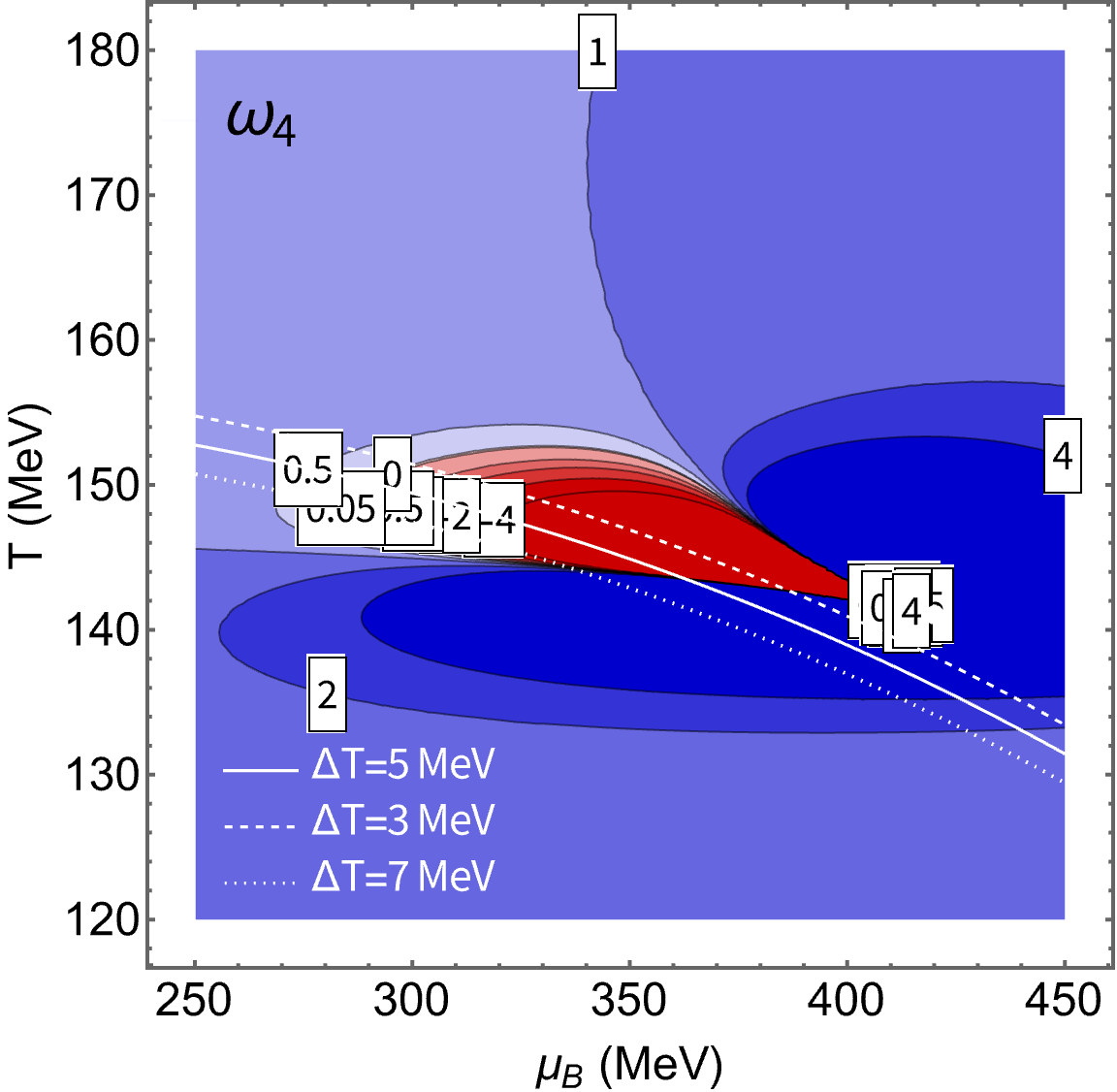}
\includegraphics[width=0.5\textwidth, height=4.5cm]{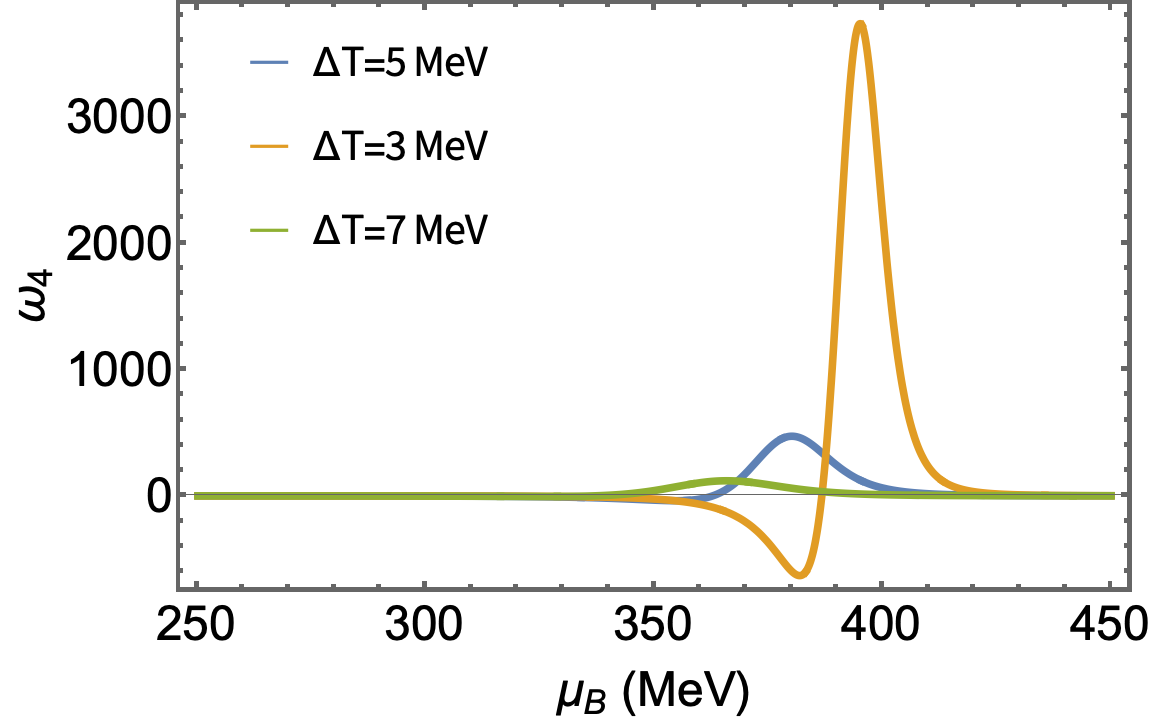}
\caption{Fourth order proton fluctuations calculated as shown in Eq. \ref{eq:omegas} in the phase diagram (left) and along freeze-out trajectories (right).}
\label{fig:w4p}       
\end{figure*}

Figure \ref{fig:w4p} shows the 4th order proton fluctuations in the phase diagram (left) and along several freeze-out trajectories (right).
The freeze-out curves are taken to be 3, 5, and 7 MeV below the critical point.
The freeze-out curve at $\Delta T = 5$ MeV is from the fit to experimental results performed in Ref. \cite{Andronic:2017pug}, while the other two are simply a shift from this established freeze-out curve to show a range of possibilities within errors.
The order of magnitude of the fluctuations is very sensitive to the path of the freeze-out curve through the phase diagram.
Taking the experimentally fitted curve at $\Delta T = 5$ MeV, the fluctuations are $\mathcal{O}(10^2)$.
However, these are still equilibrium quantities which are not to be compared with experimental results. 
A treatment such as the one in Ref. \cite{Pradeep:2022mkf} for these higher order moments is first required.
Therefore, given these large equilibrium results and what is known about the effects of freezing out critical fluctuations from Ref. \cite{Pradeep:2022mkf}, we expect the experimental results to be reduced.
Furthermore, here we have made one choice of parameters for the Ising-QCD mapping that is motivated by utilizing the available universal inputs. 
However, this remains a choice which also effects the size and shape of the fluctuations.
In a forthcoming publication, we will explore the effect of the parameters in detail.

\section{Conclusions}
Many open questions remain about the phase structure of QCD at finite temperatures and densities which are outside the current range of first-principle approaches.
I focused here on the effects of a potential critical point on the equilibrium particle fluctuations. 
By utilizing universal properties of the equation of state, we have made estimates for the equilibrium critical fluctuations of protons. 
Importantly, these equilibrium quantities are ready to in turn be used as input for estimating out-of-equilibrium fluctuations in a similar manner as was previously shown for Gaussian fluctuations in Ref. \cite{Pradeep:2022mkf} such that they can more readily be compared to experimental data.
Furthermore, given these tools to make such estimates, once the experimental data becomes available we will be able to constrain the non-universal features of this approach in order to yield the proper mapping between the QCD phase diagram and its analogous critical features from the 3D Ising model. 

\section{Acknowledgements}
The author is grateful to the conference organizers for the opportunity to present this and other work at Quark Matter 2023.
The author is supported by an Ascending Postdoctoral Scholar Fellowship from the National Science Foundation under Award No. 2138063.

\bibliography{all}

\end{document}